\documentclass[12pt]{article}

\newread\testifexists
\def\GetIfExists #1 {\immediate\openin\testifexists=#1
    \ifeof\testifexists\immediate\closein\testifexists\else
    \immediate\closein\testifexists\input #1\fi}

\usepackage{gthstyle}
\immediate\openin\testifexists=e
    \ifeof\testifexists\immediate\closein\testifexists\else
    \immediate\closein\testifexists\input e\fipsf

\def\Bbb#1{\setbox0=\hbox{$\tt #1$}  \copy0\kern-\wd0\kern .1em\copy0}

\def\bbf#1{\setbox0=\hbox{$#1$} \kern-.025em\copy0\kern-\wd0
        \kern.05em\copy0\kern-\wd0 \kern-.025em\raise.0433em\box0}

\immediate\openin\testifexists=a
    \ifeof\testifexists\immediate\closein\testifexists\else
    \immediate\closein\testifexists\input a\fimssym.def          

      \def\b{\beta}         
\def\d{\delta}      \def\D{\Delta}  
          \def\l{\lambda}     
\def\m{\mu}                     
             
     \def\s{\sigma}  
\def\t{\tau}

\def\pa{\partial} \def\ra{\rightarrow}

\def\dd{{\rm d}}

\def\fract#1#2{{\textstyle{#1\over#2}}}
\def\ffract#1#2{\raise .3 em\hbox{$\scriptstyle#1$}\kern-.25em/
                \kern-.2em\lower .2 em \hbox{$\scriptstyle#2$}}

\def\half{\fract12}  

\def\part#1#2{{\partial#1\over\partial#2}}

\def\iss{\ =\ }

\newcommand{\be}{\begin{eqnarray}}
\newcommand{\ee}{\end{eqnarray}}
\newcommand{\eqn}[1]{(\ref{#1})}
\newcommand{\nn}{\nonumber\\}

\newcommand{\bi}[1]{\begin{itemize}\item[#1]}

\newcommand{\ei}{\end{itemize}}

\newcommand{\newsec}[1]{\section{#1}\setcounter{equation}{0}}

\newcommand{\eel}[1]{\label{#1}\end{eqnarray}}\newcommand{\crl}[1]{\label{#1}\\ }

\begin{document}

\begin{titlepage}

\title{\normalsize \hfill ITP-UU-04/17  \\ \hfill SPIN-04/10
\\ \hfill {\tt hep-th/0408148}\\ \vskip 20mm \Large\bf
MINIMAL STRINGS FOR BARYONS
\thanks{Presented at the Workshop on Hadrons and Strings, \emph{Trento}, july,
2004}}

\author{Gerard 't~Hooft}
\date{\normalsize Institute for Theoretical Physics \\
Utrecht University, Leuvenlaan 4\\ 3584 CE Utrecht, the
Netherlands\medskip \\ and
\medskip \\ Spinoza Institute \\ Postbox 80.195 \\ 3508 TD
Utrecht, the Netherlands \smallskip \\ e-mail: \tt
g.thooft@phys.uu.nl \\ internet: \tt
http://www.phys.uu.nl/\~{}thooft/}

\maketitle

\begin{quotation} \noindent {\large\bf Astract } \medskip \\
A simple model is discussed in which baryons are represented as
pieces of open string connected at one common point. There are two
surprises: one is that, in the conformal gauge, the relative
lengths of the three arms cannot be kept constant, but are
dynamical variables of the theory. The second surprise (as
reported earlier by Sharov) is that, in the classical limit, the
state with the three arms of length not equal to zero is unstable
against collapse of one of the arms. After collapse, an arm cannot
bounce back into existence. The implications of this finding are
briefly discussed.
\end{quotation}

\vfill \flushleft{\today}

\end{titlepage}

\eject

\newsec{Introduction}
Mesons appear to be well approximated by an effective string
model in four dimensions, even if anomalies and lack of
super-symmetry cause the spectrum of quantum states to violate
unitarity to some extent. Presumably this is because there are no
difficulties with classical (open or closed) strings in any
number of dimensions, without any super-symmetry. String theory
can simply be seen as a crude approximation for mesons, and as
such it works reasonably well, although some observed bending of
the Regge trajectories at low energies is difficult to reproduce
in attempts at an improved treatment of the quantization in such
a regime.

\begin{figure}[h] \setcounter{figure}{0}
\begin{quotation}
 \epsfxsize=125 mm\epsfbox{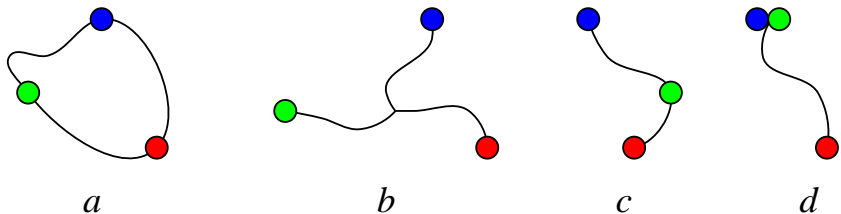}
  \caption{\small{Four possible string models for baryons.
 $a)$ Three fermions on a closed string;
 $b)$ Our starting point: quarks attached by a single string; $c)$ After
 collapse of one arm; $d)$ With two quarks in a \(\overline 3\) bound state.}}
  \label{figure1.fig}\end{quotation}
\end{figure}

It would be very desirable to have an improved effective string
model for QCD even if no refuge can be taken into any
supersymmetric deformations of the theory\cite{Polyakov}. Such a
string model should explain the approximately linear Regge
trajectories, and if we can refine it, one might imagine using it
as an alternative method to compute spectra and transition
amplitudes in mesons and baryons. Ideally, a string model would
not serve as a replacement of QCD, but as an elegant computational
approach.

The Regge trajectories for baryons seem to have the same slope as
the mesonic ones. This can easily be explained if we assume
baryons to consist of open strings that tend to have quarks at one
end and diquarks at the other. The question asked in this short
note is how a classical string model for baryons should be
handled\cite{Artru}\cite{Sh9809}. In the literature, there appears
to be a preference for the \(\D\) model\cite{Sh9808}, consisting
of a closed string with three fermion-like objects attached to it,
see Fig. \ref{figure1.fig}$a$. As closed strings can be handled
using existing techniques, this is a natual thing to try.

However, from a physical point of view, the Y configuration,
sketched in Fig. \ref{figure1.fig}$b$ appears to be a better
representation of the baryons. After all, given three quarks at
fixed positions, the Y shape takes less energy. Furthermore, if
two of these quarks stay close together, they behave as a diquark
and the Regge spectrum (with the \emph{same} slope as the mesons)
would be readily explained. However, there turns out to be a good
reason why the Y shape is dismissed\cite{Sh0004}; we will discuss
this, and then ask again whether ``\(\D\)" is to be preferred.

We considered the exercise of solving the classical string
equations for this configuration. After we did this exercise, we
found that it has been discussed already by Sharov\cite{Sh0001}.
The question will be whether we agree with his conclusions.

\newsec{The three arms}
The three arms are described by the coordinate functions
\(X^{\m,i}(\s,\,\t)\), where the index \(i\) is a label for the
arms: \(i=1,2,\) or \(3\). At the points \(\s=0\), the three arms
are connected: \be X^{\m,1}(0,\,\t)= X^{\m,2}(0,\,\t)=
X^{\m,3}(0,\,\t)\ . \eel{boundary0} \(\s\) is a coordinate along
the three arms. It takes values on the segment
\([\,0,\,L^i(\t)\,]\), where as yet we keep the lengths
\(L^i(\t)\) unspecified.
\begin{figure}[h]\begin{quotation}
 \epsfxsize=70 mm\epsfbox{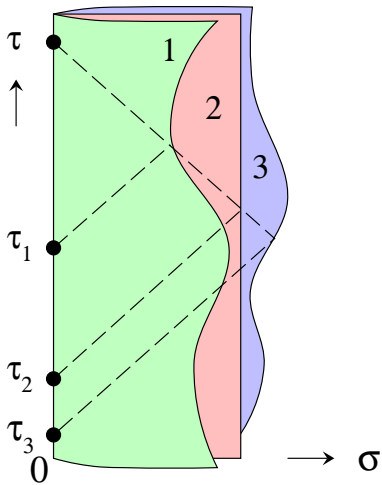}
  \caption{\small{The three world sheets. Their relative lengths are dynamical
  variables.   One of the three may be kept fixed.   Dashed lines: bouncing waves
  have differing arrival times.}}
  \label{figure2.fig}\end{quotation}
\end{figure}
The Nambu action is\be
S=-\sum_{i=1}^3\int\dd\t\int_0^{L^i(\t)}\dd\s\, \sqrt{(\pa_\s
X^{\m,i} \cdot\pa_\t X^{\m,i})^2-(\pa_\s X^{\m,i})^2(\pa_\t
X^{\l,i})^2}\,\Big)\ . \eel{Nambu} Here, as in the expressions
that follow, we assume the usual summation convention for the
Lorentz indices \(\m,\ \l,\ \dots\), but not for the branch
indices \(i\), where summation, if intended, will always be
indicated explicitly. For simplicity, the masses of the quarks at
the end points are neglected.

Assuming, in each of the three arms, the classical gauge condition
\be \pa_\s X^{\m,i}\cdot\pa_\t X^{\m,i}=0\ ; \crl{gaugecond1}
(\pa_\s X^{\m,i})^2=-(\pa_\t X^{\m,i})^2\ , \eel{gaugecond2}
Eq.~\eqn{Nambu} takes the form \be
S=\sum_{i=1}^3\int\dd\t\int_0^{L^i(\t)}\dd\s\Big(\half(\pa_\t
X^{\m,i})^2-\half(\pa_\s X^{\m,i})^2\Big)\ . \eel{action}

\newsec{Boundary conditions} The boundary condition at the origin
can be enforced by two Lagrange multipliers: we add to the action
\eqn{action}
\be\int\dd\t(\l_1^\m(\t)(X^{\m,1}(0,\t)-X^{\m,3}(0,\t))+
\l_2^\m(\t)(X^{\m,2}(0,\t)-X^{\m,3}(0,\t))\ . \eel{Lagrange} Now,
considering an infinitesimal variation of \(X^{\m,i}(\s,\,\t)\),
we find \be\d S&=&\sum_i\bigg(\int\dd\t\dd\s\,\d
X^{\m,i}(\pa_\s^2-\pa_\t^2)X^{\m,i}\ +\nn &&\int\dd\t\Big(\d
X^{\m,i}(0,\,\t) (\pa_\s X^{\m,i}(0,\,\t)+\l^\m_i)-\d
X^{\m,i}(L^i,\,\t)\pa_\s X^{\m,i}(L^i(\t),\,\t)\Big)\bigg)\ ,
\eel{vary} where \(\l_3=-\l_1-\l_2\,\). Thus, we see that the
usual Neumann boundary condition holds at the three end points
\(\s=L^i\), while at the origin we have a Neumann boundary
condition for the sum \(\sum_{i=1}^3 X^{\m,i}\) and a Dirichlet
condition for the differences: \be \s=0\ :&\quad&
\pa_\s(X^{\m,1}+X^{\m,2}+X^{\m,3})=0\ ; \nn &&
X^{\m,1}-X^{\m,3}=X^{\m,2}-X^{\m,3}=0\ ,\nn \s=L^i(\t)\ :&\quad&
\pa_\s X^{\m,i}=0\ ,\quad \forall i\ . \eel{bc}
\newsec{Equations of motion} From Eq.~\eqn{vary}, we read off the
usual equations of motion implying that we have the sum of left-
and right-going waves on all three branches: \be
X^{\m,i}=X^{\m,i}_L(\s+\t)+X^{\m,i}_R(\t-\s)\ . \eel{LRmodes} The
boundary conditions \eqn{bc} now read \be
X^{\m,i}_L=X^{\m,i}_R\qquad&\mathrm{at}&\ \s=L^i(\t)\ ,\quad
\mathrm{and}\crl{endboundaries} \matrix{
 \pa_\t X^{\m,1}_R&=&\pa_\t(\fract23 X_L^{\m,2}+\fract23
X_L^{\m,3}-\fract13 X_L^{\m,1}) \cr
 \pa_\t X^{\m,2}_R&=&\pa_\t(\fract23 X_L^{\m,3}+\fract23
X_L^{\m,1}-\fract13 X_L^{\m,2}) \cr
 \pa_\t X^{\m,3}_R&=&\pa_\t(\fract23 X_L^{\m,1}+\fract23
X_L^{\m,2}-\fract13 X_L^{\m,3}) } \ \Bigg\}\ &\mathrm{at}& \
\s=0\ . \eel{midboundary}

In addition to these, we have the constraints \eqn{gaugecond1}
and \eqn{gaugecond2}, now taking the form \be\sum_\m(\pa_\t
X^{\m,i}_L)^2=\sum_\m(\pa_\t X^{\m,i}_R)^2=0\ . \eel{constraints}
These constraints should be compatible with the boundary
conditions. The conditions at the three end points,
Eqs.~\eqn{endboundaries} give nothing extra, but the equations at
the origin, Eqs.~\eqn{midboundary}, require that, at that point,
also \be \pa_\t X^{\m,1}_L\,\pa_\t X^{\m,2}_L\iss\pa_\t
X^{\m,2}_L\,\pa_\t X^{\m,3}_L\iss\pa_\t X^{\m,3}_L\,\pa_\t
X^{\m,1}_L\ ,\eel{cross} and similarly for the right-going modes.
From the point of view of causality, it might seem odd that there
is a condition that the newly arriving modes should obey.
However, it can easily be imposed by a relative rescaling of the
coordinates \(\,\s+\t\,\) for the left-going modes in the three
arms: \be \pa_\t X^{\m,i}_L(\t)\iss {\dd\t_i\over\dd\t}\
\pa_{\t_i} X^{\m,i}_L(\t_i)\ . \eel{rescale} Indeed, as we shall
now demonstrate, the requirement \eqn{cross} fixes the functions
\(L^i(\t)\).

Taking the fact that \(X^{\m,i}_L\) only depends on \(\,\t+\s\,\)
and \(X^{\m,i}_R\) only on \(\,\t-\s\,\), the boundary condition
\eqn{endboundaries} implies the existence of functions
\(\t_i(\t)<\t\) such that \be X^{\m,i}_L(\t)=X^{\m,i}_R(\t_i)\ ,
\eel{periods} where \(\t_i(\t)\) are the solutions of \be
\t-\t_i\iss 2L^i(\t'_i)\ ,\qquad\t'_i\equiv{\t+\t_i\over 2}\ .
\eel{delays} These are just the three departure times, from the
center, of each of the three returning waves, see
Fig.~\ref{figure2.fig}. Now let \be Q^{\m,i}(\t)&\equiv&{\pa
X^{\m,i}_R(\t)\over\pa\t}\ ;\crl{Qdef}
R^i(\t)&\equiv&-\,Q^{\m,j}(\t_j)\,Q^{\m,k}(\t_k)\ ,\qquad
i,j,k=1,2,3 \ \mathit{cyclic,} \eel{R123} (i,j,k being any even
permutation of 1,2,3). The minus sign in Eq.~\eqn{R123} is chosen
because \(Q^{\m,i}\,Q^{\m,j}=\vec Q^i\cdot\vec Q^j-|Q^i||Q^j|\le
0\,\), since \((Q^{\m,i})^2=0\,\). Then, using \eqn{rescale},
condition \eqn{cross} implies \be {1\over
R^1(\t)}{\dd\t_1\over\dd\t}\ =\ {1\over
R^2(\t)}{\dd\t_2\over\dd\t}\ =\ {1\over
R^3(\t)}{\dd\t_3\over\dd\t} \ . \eel{tauequ} Together with a gauge
condition for the overall time parameter \(\t\), for instance, \be
\fract13\sum_i\t_i= \t-1\ ;\qquad\sum_i{\dd\t_i\over\dd\t}=3\ ,
\eel{taugauge} these form complete differential equations that
determine the functions \(\t_i(\t)\), and thus also the functions
\(L^i(\t)\), through Eqs.~\eqn{delays}: \be {\dd\t_i\over\dd\t}={3
R^i\over R^1 + R^2 +  R^3 }\ . \eel{taui} The complete update of
the functions \(Q^{\m,i}(\t)\) at the central point then reads:
\be {Q^{\m,i}}(\t)\iss {1\over \sum R}\Big(2R^j(\t)Q^{\m,j}(\t_j)+
2R^k(\t)Q^{\m,k}(\t_k)-R^i(\t)Q^{\m,i}(\t_i)\Big)\ ,&&\nn
i,j,k=1,2,3 \ \mathit{cyclic.}&& \eel{Qiupdate}   Using
Eqs.~\eqn{R123}, one verifies that indeed the constraint
\((Q^{\m,i}(\t))^2=0\,\) continues to be respected.
\newsec{An instability}
In the special case when all \(R^i\) happen to be constant and
equal, these equations can be solved. The three arms then have
equal lengths, in the sense that waves running across the three
arms bounce back in exactly the same time, which is constant.
However, one readily convinces oneself that solutions with
unequal arms should also exist. For instance, take an initial
state described by functions \(Q^{\m,i}(\t)\) consisting of
superpositions of theta functions in \(\t\), such that the
condition \be \sum_\m(Q^{\m,i})^2=0\eel{Qconstr} is always
satisfied. One can then solve the equations stepwise for
successive intervals in \(\tau\).

\begin{figure}[h]\begin{quotation}
 \epsfxsize=120 mm\epsfbox{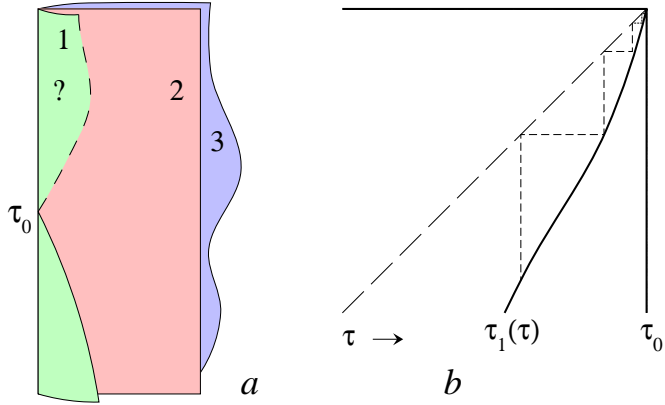}
  \caption{\small{$a)$  A contracting arm. $b)$ Indefinite series of `updates' of
the data along the contracting arm, due to bouncing waves.}}
  \label{figure3.fig}\end{quotation}\end{figure}

There is, however, a problem that needs to be addressed: what
happens if one of the arms tends to vanish? This is for instance
the case if, during some period, \(\dd\t_1/\dd\t>1\) until
\(\t-\t_1\ra0\), see Fig.~\ref{figure3.fig}$a$. Do we need a new
boundary condition addressing this situation?  Indeed, it appears
that, at least in the classical (unquantized) theory, an extra
boundary condition will be needed. Our argument goes as follows.

If, at some value(s) of \(\t\), one arm length parameter, say
\(L^1(\t)\,\), tends to zero, then there will be waves running up
and down this arm increasingly frequently, see
Fig.~\ref{figure3.fig}$b$. The values for \(Q^{\m,i}\) for the two
long arms, \(i=2,3\), will not vary rapidly during this small
period (this is because these incoming waves were last updated at
the times \(\t_2\) and \(\t_3\), both long before \(\t\)). Let us
make an educated guess about what will happen, by keeping these
\(Q\) vectors constant. In contrast, \(Q^{\m,1}(\t)\) will be
updated frequently, so, let us consider this sequence. We have \be
{Q^{\m,1}}(\t)&=& {1\over \sum_i R^i}\Big(2R^2(\t)Q^{\m,2}(\t_2)+
2R^3(\t)Q^{\m,3}(\t_3)-R^1(\t)Q^{\m,1}(\t_1)\Big)\ ,
\crl{Q1update} && Q^{\m,2}(\t_2)\quad\mathrm{and}\quad
Q^{\m,3}(\t_3) \quad \mathit{constant}. \eel{Q23update} One finds
that the parameters \(R^i\) are updated as follows: \be R^1&\ra&
R^1\equiv R\quad\mathrm{(constant)}\ ;\nn R^2&\ra& {R^1\,R^2\over
R^1+R^2+R^3}\ ;\nn R^3&\ra& {R^1\,R^3\over R^1+R^2+R^3}\ .
\eel{Rupdate} This implies that \(R^2\) and \(R^3\) continue to
decrease, until \(R^1\) dominates. Thus, \be
{\dd\t_1\over\dd\t}\ra 3\ ,\qquad {\dd\t_2\over\dd\t}\ra 0\
,\qquad {\dd\t_3\over\dd\t}\ra 0\ . \eel{taulimit} Since we
started with \(\t_1<\t\), we find that \(\t_1\ra\t\,\), rapidly.
In fact, we found that, continuing along this line, assuming
\(Q^{\m,2}\) and \(Q^{\m,3}\) to stay constant, the calculation
can be done exactly. After \(n\) iterations, the vector
\(Q^{\m,1,\,(n)}\) becomes \be Q^{\m,1\,(n)}={1\over
n}\Big(Q_0^\m(-1)^n+\b Q^{\m,2}+(1-\b)Q^{\m,3}\Big)\ ,
\eel{Qiterate} where \(Q_0^\m\) is a fixed vector, and \(\b\) is a
fixed coefficient, such that \be Q_0^\m\,Q^{\m,2}=
Q_0^\m\,Q^{\m,3}=0\ ;\qquad (Q_0^\m)^2=R\b(1-\b)\ ;\qquad
Q^{\m,2}\,Q^{\m,3}=-R\ . \eel{iterationconsts} We also have \be
R^{2\,(n)}={\b\over n}R\ ;\qquad R^{3\,(n)}={1-\b\over n}R\ .
\eel{R23iterate} In fact, \(Q_0^\m\) and \(\b\) are only constant
while following a bouncing wave, but they cannot be constant along
the entire short arm. This is because of the alternating sign in
Eq.~\eqn{Qiterate}. Since \(Q_0^\m\) is spacelike and orthogonal
to \(Q^{\m,2}\) and \(Q^{\m,3}\), it must oscillate through zero.
At such points, either \(\b\) or \(1-\b\) must have quadratic
zeros. Typically, one of these will oscillate like
\(\sin^2(\pi\s/2L^1)\,\). Thus, our solution consists of
partly-periodic functions, in the sense that over periods lasting
approximately \(\ln 3\) in the parameter \(\ln(\t_0-\t)\), the
functions are updated using Eqs.~\eqn{Qiterate}, where the
auxiliary functions \eqn{iterationconsts} are exactly periodic.

We have no indication that small perturbances in the initial
conditions will have drastic effects on these solutions, so that
indeed generic solutions exist in which one arms rapidly shrinks
to zero. What happens after such an event? At first sight, it
seems reasonable to postulate some sort of bounce. By
time-reversal symmetry, one might expect this arm to come back
into existence. However, closer inspection makes such a
`solution' quite unlikely, or even impossible. Our observation is
the following. Shortly before our shrinking event, the short arm
has been violently oscillating, or rotating, around the central
point. In doing so, the rapidly oscillating function \(Q^{\m,1}\)
has been emitting high-frequency waves into the two long arms.
These frequencies will always be much higher than the frequencies
of oscillations entering from the long arms. The time-reversal of
this configuration is hard to reconcile with causality
requirements. Thus, the energy of the short arm has been
dissipating in the form of high-frequency modes in the long arms,
and the dissipated energy will be hard to recover.

\newsec{Conclusion}
We did not study possible quantization procedures for this model
in any detail. Extreme non-linearity at the connection point
probably makes this impossible. Qualitatively, as is well-known,
one expects Regge trajectories that are similar --- and have the
same slope --- as the mesonic ones, since at any given energy, the
highest angular momentum states will be achieved when one arm
vanishes. Now, in our analysis, we find that, as soon as a
classical string picture is adopted for baryonic states, at least
one of the three arms will soon disappear, shedding its energy
into the excitation modes of the two other arms, see
Fig.~\ref{figure1.fig}$c$.

This is somewhat counter-intuitive. One might have thought that
equipartition should take place: all corners of phase space should
eventually be occupied with equal probability. The answer to this
is, of course, that phase space is infinite, so that equipartition
is impossible. It is the old problem of statistical physics before
the advent of Quantum Mechanics: there is an infinite amount of
phase space in the high frequency domain. Since physical baryons
are quantum mechanical objects, we expect an effective cut-off at
high frequencies, simply because of energy quantization. This does
mean, however, that in the high energy domain, the majority of
baryonic states will have these high-energy modes excited. If, due
to some electro-weak interaction, or possibly due to a gluon
hitting a quark, a baryonic state is created with three quarks
energetically moving in different directions, we expect first the
Y shape to form, but then the most likely baryonic excitation that
is reached is one with a single open string connecting the three
quarks.

We found that the classically stable configuration has a single
open string with two quarks at the end points and one quark moving
around on the string. However, because of an attraction between
two quarks into a \(\overline 3\) bound state (as opposed to the
6), one expects quantum effects eventually to favor the
configuration of one quark at one end and a di-quark at the other
end of a single open string, see Fig.~\ref{figure1.fig}$d$. I
still do not see a strong case in favor of the \(\D\)
configuration. The latter seems to be closer related to a
baryon-glueball bound state that will probably have a rather low
production cross section.

Finally,  Y shaped string configurations have also emerged in
string theory\cite{Witten}, but only by viewing the connection
point as a \(D\)-brane, which as such must be handled in the
classical approximation, so that most of the mass of the system is
concentrated there. The dynamical properies of such configurations
will again be different from what we studied here.

\end{document}

hep-ph/9812527